\documentclass[12pt]{iopart}

\usepackage{appendix}
\usepackage{SIunits}
\usepackage[numbers,compress]{natbib} 
\usepackage{sistyle}
\usepackage[utf8x]{inputenc}
\usepackage{graphicx}   
\usepackage{dcolumn}
\usepackage{bm}
\usepackage{hyperref}
\usepackage{epstopdf}
\usepackage{lipsum}
\usepackage{braket}
\DeclareGraphicsExtensions{.pdf,.png,.jpg,.jpeg,.mps}
\usepackage{pgf}
\usepackage{tikz}
\usepackage{xr}

\newcommand{\figref}[1]{Fig.~\ref{#1}}

\newcommand{\secref}[1]{Sec.~\ref{#1}}
\newcommand{\eqnref}[1]{Eq.~\ref{#1}}

\renewcommand{\figurename}{Fig.}
\newcommand{\akuts}[1]{``#1''}

\begin{document}
\title {Steering between Level Repulsion and Attraction: {\color{black} Broad tunability of Two-Port} Driven Cavity Magnon-Polaritons}

\author{Isabella Boventer}
\address{Institute of Physics, Johannes Gutenberg University Mainz, 55128 Mainz, Germany}
\address{Unit\'{e} Mixte de Physique CNRS, Thales, University Paris-Sud, Universit\'{e}  Paris-Saclay, 91767 Palaiseau, France}
\author{Mathias Kl\"aui}
\address{Institute of Physics, Johannes Gutenberg University Mainz, 55099 Mainz, Germany}
\address{Graduate School of Excellen Materials Science in Mainz, Staudinger Weg 7, 55128 Mainz, Germany}


\author{Rair Mac{\^e}do}
\address{James Watt School of Engineering, Electronics \& Nanoscale Engineering Division, University of Glasgow, Glasgow G12 8QQ, United Kingdom}

\author{Martin Weides}
\ead{Martin.Weides@glasgow.ac.uk}
\address{James Watt School of Engineering, Electronics \& Nanoscale Engineering Division, University of Glasgow, Glasgow G12 8QQ, United Kingdom}


\date{\today}

\begin{abstract}
Cavity-magnon polaritons (CMPs) are the associated quasiparticles of the hybridization between cavity photons and magnons in a magnetic sample placed in a microwave resonator. 
In the strong coupling regime, where the macroscopic coupling strength exceeds the individual dissipation, there is a coherent exchange of information.  
This renders CMPs as promising candidates for future applications such as in information processing. 
Recent advances on the study of the CMP now allow not only for creation of CMPs on demand, but also for tuning of the coupling strength - this can be thought of as enhancing or suppressing of information exchange. 
Here, we go beyond standard single-port driven CMPs and employ a two-port driven CMP. 
We control the coupling strength by the relative phase $\phi$ and amplitude field ratio $\delta_0$ between both ports. 
Specifically, we derive a new expression from Input-Output theory for the study of the two-port driven CMP and discuss the implications on the coupling strength. 
{\color{black}Furthermore, we examine intermediate cases where the relative phase is tuned between its maximal and minimal value and, in particular, the high $\delta_0$ regime, which has not been yet explored.}   
\end{abstract}
\vspace{2pc}

\maketitle
\section{Introduction}
The phenomena of a (strong) coupling of magnons -- the associated quanta of collective spin wave excitations -- to microwave cavity photons, resulting in cavity magnon-polaritons (CMPs) has been the subject of numerous works in the past few years \cite{Soykal2010,Huebl2013,Goryachev2014,Tabuchi_2014, Zhang_2014,Zhang2015,Harder2018,Quirion2019, Pfirrmann2019}. The ability to couple magnons to different physical systems, through magneto-optical \cite{Sharma2017,Kusminskiy2016,Graf2018} to optical, or by magnetostrictive interaction to mechanical \cite{Tang2016} and cavity photons simultaneously makes CMPs highly interesting for various applications \cite{Quirion2019}. For instance, it allows for a bidirectional conversion of microwaves to optical light \cite{Hisatomi2016}, or coupling magnons with superconducting circuits, i.e. qubits \cite{Tabuchi2015}. The context of these studies varies from purely classical  \cite{Sharma2017, Hisatomi2016,Haigh2016,Yao2017,Rao2019} to quantum based approaches \cite{Tabuchi2015,Wang2019}. 
For a strongly coupled cavity-magnon system where the coupling strength exceeds the individual dissipation from each subsystem at resonance, that is $\omega_c=\omega_m \equiv \omega_0$, the cavity photon ($\omega_c$) and magnon states ($\omega_m$) hybridize. As a result of the simultaneous coupling of N contributing spins of the magnonic sample, one observes the opening of a frequency gap $\Delta \omega=2g_{\mathrm{eff}}=2g_{0}\sqrt{N}$ due to level repulsion \cite{Tabuchi_2014}.  Here, $g_0$ denotes the single spin and $g_{\mathrm{eff}}$ denotes the effective macroscopic coupling strength in the dispersion spectrum. 
{\color{black} It is worth noting that, the single spin coupling strength $g_0=\frac{\eta \gamma}{2}\sqrt{\frac{\mu_0\hbar\omega_0}{2V_{\mathrm{mode}}}}$ does not depend on the photon number.  Rather, $g_0$ is determined by the photon and magnon mode overlap $\eta$, the resonance frequency $\omega_0$ and mode volume $V_{\mathrm{mode}}$ of the chosen cavity resonator mode.}  
The observation of such an avoided crossing (anti-crossing) is a characteristic feature of cavity magnon-polaritons (CMPs) and it enables the study of properties of said systems \cite{Harder2018}. 
\newline
However, in most of these works, being able to control the coupling constant is imperative; whether the ultimate goal is to achieve stronger coupling or to control the actual state of coupling \cite{Zhang2017a}. 
While most of the above mentioned initial studies have concentrated in the case of level repulsion which leads to said avoided level crossing -- also known as Rabi splitting \cite{Walls2008} -- more recently, another phenomenon has emerged which is called level attraction \cite{Bernier2018,Grigoryan2019,Harder2018a,Grigoryan2018,proskurin2019}. In order to achieve enter the regime of level attraction, several approaches have been employed so far. The most simplistic one, perhaps, is moving the magnetic sample to different positions within the 3-dimensional microwave resonator \cite{Harder2018a} or even a 2-dimensional one \cite{Bhoi2019}. In most of these experiments, however, a microwave signal from an external source was coupled into the resonator and thereby directly driving the cavity photons at a certain cavity resonance frequency $\omega_c$. In such setups, a magnetic sample was placed into an antinode of the time varying magnetic field from the chosen cavity resonator mode resulting from alternating currents (AC). These AC fields would then drive ferromagnetic resonances in the magnetic sample, i.e. it would excite magnons resonating at a frequency $\omega_m$ \cite{Soykal2010}.

In a recent work, we have shown a way to access the regime of level attraction by the addition of a second external microwave input and by externally controlling the relative phase $\phi$ and internal amplitude ratio $\delta_0$ of the AC magnetic fields within the resonator {\color{black} via tuning the relative input amplitude at each microwave input. } \cite{Boventer2019}. 
By tuning the relative phase to $\phi=\pi$ and setting $\delta_0=1$, we observed a full closure of the anticrossing gap which we also call level merging. {\color {black}Experimentally, the relative phase shift is realized by the addition of a mechanically tunable phase shifter in the signal path to the magnon port (c.f. \figref{Setup})}. If the phase is kept fixed and $\delta_0>1$, we enter the regime of level attraction. 
In this work, we study  the conditions under which this coupling might happen in detail. We further study intermediate phases where level repulsion and attraction are both present {\color{black}which has not been observed previously as well as} the impacts of a higher value of $\delta_0$ on our system. We focus on the coupling of cavity photons to magnons in the Kittel mode, which is a special instance of a magnetostatic mode with wavevector $\mathbf{k}=0$. The Kittel mode denotes the uniform precession for all spins and has a dispersion $\omega_m=\gamma \mathbf{H}_{0}$, where $\gamma$ is the gyromagnetic ratio and $\mathbf{H}_{0}$ is a static magnetic field externally applied \cite{AG2000}. As it typically shows the highest coupling strength to the cavity photons, numerous experiments studied CMPs via the coupling to the Kittel mode \cite{Huebl2013, Tabuchi_2014,Zhang_2014,Harder_2016}.  

The experimental setup is described in Sec.  \ref{Sec:Setup}, followed by the theory detailed in Sec. \ref{Sec:The} and the experimental results and discussions are given in Sec. \ref{Sec:Res}.

\section{Experimental Setup}\label{Sec:Setup}

\begin{figure}
\centering
\includegraphics[scale=0.4]{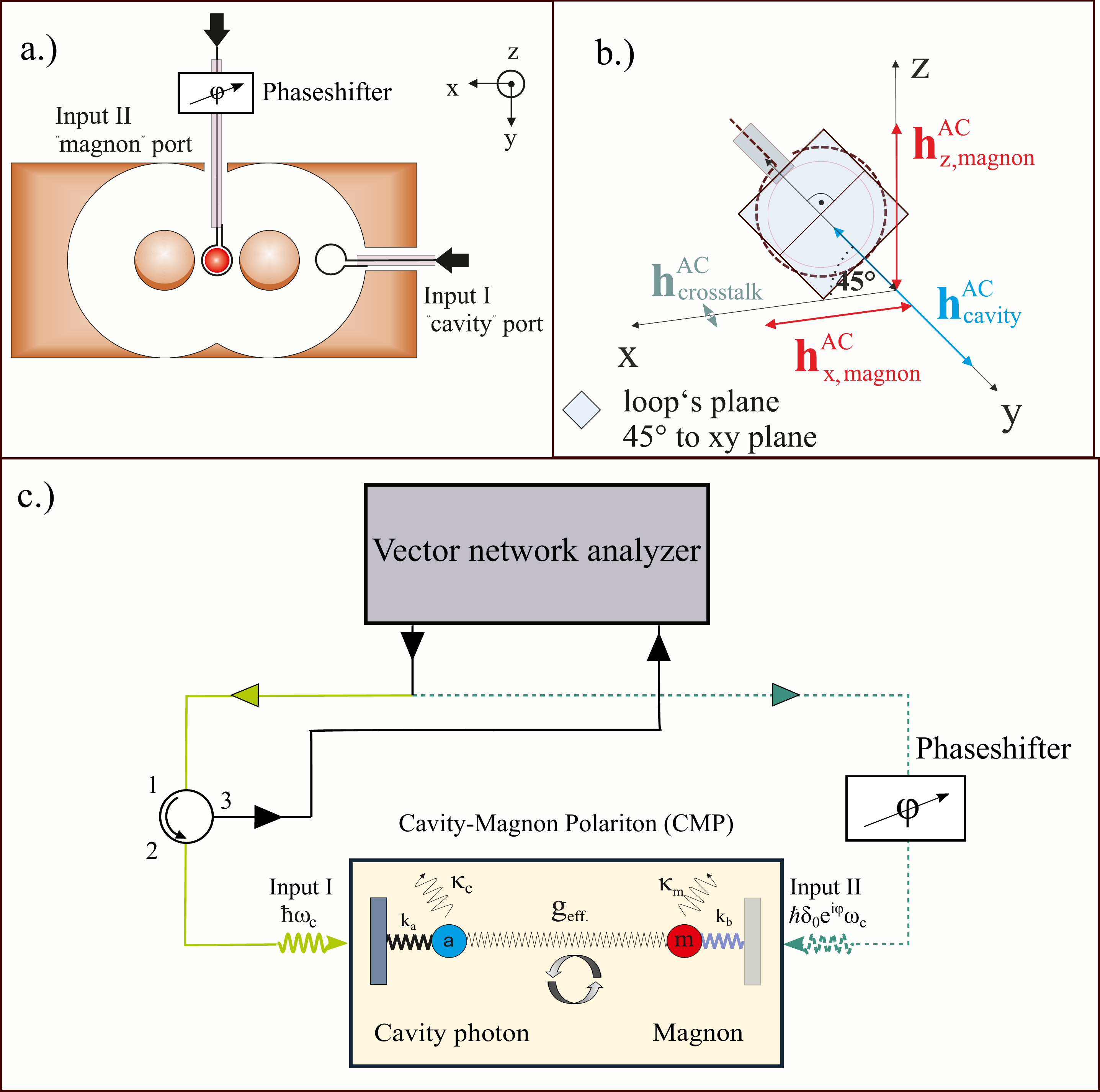}
\caption{Overview over the implementation of the two ports for the coupling strength control of the CMP. a.) Topview: Position of both inputs {\color{black}including the mechanically tunable phaseshifter}, where the microwave signal is inductively coupled by a single winded metallic loop into the cavity resonator. b.) Relative orientation of the magnon port's coupling loop around an sphere made of Yttrium-Iron-Garnet (YIG) and alignment of the intracavity AC magnetic fields. c.) Schematics of the experimental setup for a single port driven (solid lines) and a two-port driven (dotted green line) CMPs measurement in reflection from the cavity port {\color{black}and the phase shifter along the path to the magnon port}. The CMP is illustrated as a system of two harmonic oscillators coupled by springs. Here, the spring constants represent the effective coupling strength. Whilst the spring constants $k_a$ and $k_b$ give the coupling efficiency to the microwave feedline, the internal losses from each constituent are given by $\kappa_c$ and $\kappa_m$ }
\label{Setup}
\end{figure}

Up until now, there have been several different and well established methods to probe the coupling between cavity and magnons experimentally where one of the most common ones is microwave spectroscopy. In this, the system's transmission or (and) reflection parameters are recorded \cite{Harder_2016}. Another method is electrical detection employing a voltage generated from a combination from spin pumping and the inverse spin Hall effect \cite{Cao2015,Bai2015}. {\color{black}Magnon induced Brillouin light scattering  has  also been recently employed within the emerging field of cavity optomagnonics.}
\cite{Osada2016}
\newline
For our two-port driven CMP experiment, we employed microwave resonator spectroscopy and modified a previous single port driven setup \cite{Boventer2019,Boventer2018}. 
In our experimental setup, we employ a reentrant cavity resonator with resonance frequency $\omega_c/2\pi=6.5\,\mathrm{GHz}$ and insert a commercially bought sphere  ($d=0.2\,\mathrm{mm}$, \cite{ferrisphere}) made of Yttrium-Iron-Garnet (YIG) into the antinode of the resonator' s AC magnetic field \cite{Goryachev2014,Boventer2018}.  
Accordingly, \figref{Setup} gives a detailed overview of the position of the two microwave inputs [topview, a.)], the relative orientation of the single winded metallic loop which constitutes the second input, called magnon port, in combination with the AC magnetic fields at the sample's position [b.)] and the complete experimental apparatus [c.)]. The Vector Network Analyzer (VNA) serves as the single microwave source of the system as illustrated in \figref{Setup}. It is split using a power divider into two signal paths.  
As can be inferred from \figref{Setup} b.), the magnon port is tilted by $45^\circ$ to the cavity resonator's xy-plane. We found experimentally, that this angle {\color{black} not only} gives the best compromise between minimal crosstalk and spatial limitations of our experimental setup, {\color{black} but it is also crucial for the observation of level attraction in our two-port driven approach (c.f. discussion and comparison to related works in Sec. 4)}. 
The non-zero angle out of the xy-plane results in two AC magnetic field components (red), i.e. $h_{\mathrm{z,\, magnon}}^{\mathrm{AC}}$ and $h_{\mathrm{x,\, magnon}}^{\mathrm{AC}}$. There, $h_{\mathrm{z,\, magnon}}^{\mathrm{AC}}$, is parallel to the direction of the external, static magnetic field $\mathbf{H}_{\mathrm{ext}}=(0,0,H_{\mathrm{ext}})$ and, hence, does not drive the magnons. 
However, $h_{\mathrm{x,\, magnon}}^{\mathrm{AC}}$ is oriented such that it also drives ferromagnetic resonance but does not directly couple to the cavity photon field (blue) because $h_{\mathrm{x,\, magnon}}^{\mathrm{AC}} \bot h_{\mathrm{cavity}}^{\mathrm{AC}}$. Thus, both inputs can be considered to act independently on the magnons, once indirectly by the first input, also called cavity port, via the coupling at resonance and directly by the second input, i.e. the magnon port. 
Experimentally, there is a suppressed but non-zero residual direct coupling to the cavity photons by a small component parallel to $h_{\mathrm{cavity}}^{\mathrm{AC}}$. This crosstalk may be, for instance, caused by another small tilt of the coupling loop along the xz-plane.  
Specifically, in the experiment, we measure at the cavity port and record the reflection parameter $S_{11}(\omega)$ at the second port of a vector network analyzer (c.f. \figref{Setup} c.)). There, we sketch the experimental setup for both a single tone CMP measured in reflection mode at the cavity port (the dashed parts are then not to be included) and a two-port driven CMP. The latter is depicted by the green dotted line.

In \figref{Setup} c.), we illustrate the CMP by a system of two coupled harmonic oscillators with individual loss parameters $\kappa_c$ for the cavity photons and $\kappa_m$ for the magnons and corresponding coupling losses due to the coupling to the external microwave feedline (c.f. Ref.\cite{Harder2018}). 
Now, such an introduction of a second input in the the experimental setup for the study of the CMP, requires the modification of the standard reflection parameter $S_{\mathrm{11}}(\omega)$. This is discussed in the next section.  
\section{Theoretical Background}\label{Sec:The}

Here, we derive an expression to model a two-port cavity magnon-polariton spectroscopic experiment measured in reflection.
However, in order to show clearly the implications of a second input,we find it necessary to introduce the concepts and main assumptions for the study of a simple single port driven CMP first.  

\subsection{Spectroscopy with one port}
An experimental setup such as the one shown in Fig.~\ref{Setup} c.) -- ignoring the dotted green line which its implications will be later discussed -- can be used to conduct measurements for single port driven CMPs.  
Such CMPs can be modeled by employing the Input-Output formalism within the framework of the Hamiltonian approach \cite{Walls2008}. In general, the Hamiltonian describing the whole system can be written as: 
\begin{equation}
\mathcal{H}=\mathcal{H}_{\mathrm{sys}}+\mathcal{H}_{\mathrm{bath}}+\mathcal{H}_{\mathrm{int}},
\label{SingleToneH}
\end{equation} 
where $\mathcal{H}_{\mathrm{sys}}=\hbar \omega_c aa^\dagger+\omega_m mm^\dagger+\hbar g_{\mathrm{eff}}(a^\dagger m+m^\dagger a)$ which is also known as the Tavis-Cummings Hamiltonian for an N particle two-level system \cite{Walls2008}. Here,  $\mathcal{H}_{\mathrm{sys}}$ refers to the intracavity interactions such as the coupling between cavity photon and magnon where  $a,a^\dagger, m,m^\dagger, g_{\mathrm{eff}}$ denote the photon destruction and creation operators of the cavity photons, the magnons and the effective macroscopic coupling strength, respectively. $\mathcal{H}_{\mathrm{bath}}$ describes the coupling to the external environment, i.e the bath; and $\mathcal{H}_{\mathrm{int}}$ is the interaction between the external field modes and the internal cavity photons.
In the most simplistic case, we assume that there is no direct coupling of the intracavity system with the environment. Accordingly, we consider the Hermitian form of this Hamiltonian. 
We can then write equations of motion (EOM) for both the cavity photons ($a,a^\dagger$) and the magnons ($m,m^\dagger$), which include damping and diffusion as:
\begin{equation}
 \frac{dm}{dt} = -\frac{i}{\hbar}[m,\mathcal{H}_{\mathrm{sys}}]-\kappa_m \cdot m,\,\\
\frac{da}{dt} =-\frac{i}{\hbar}[a,\mathcal{H}_{\mathrm{sys}}]-\kappa_c \cdot a +\sqrt{2\kappa_e}b_{\mathrm{in}}(t).
\end{equation}

These expressions can then be combined in order to derive reflection $S_{11}(\omega)$ or transmission $S_{21}(\omega)$ parameters from Input-Output theory. However, these steps are familiar from Refs. \cite{Walls2008,Harder2018a} for reflection and from Ref. \cite{Tabuchi_2014} for transmission. Thus, we only summarize the basic assumptions in order to obtain the final equations. These are:  
\begin{itemize}
\item[1.] The magnons are not coupled to the external bath, but solely to the cavity photons. 
\item[2.] The photons are coupled to the external bath which represents the input microwave field from the cavity port.   
\item[3.] The following Input-Output relation between the signal entering and leaving the cavity resonator is utilised \cite{Walls2008}: $b_{\mathrm{out}}(\omega)+b_{\mathrm{in}}(\omega)=\sqrt{2\kappa_{e,i}}a(\omega)$, where $b_{\mathrm{out}}(\omega)$ and $b_{in}(\omega)$ denote the output and input from the microwave feedline to the cavity resonator port, respectively, and $a(\omega)$ is the internal cavity photon field. 
\end{itemize}
The EOMs are then solved and, by means of a Fourier transformation,  expressed as functions of the frequencies $\omega$. These yield
\begin{equation}
S_{\mathrm{11}}(\omega)=-1 + \frac{2\kappa_e}{i(\omega_c-\omega)+\kappa_c+\frac{g_{\mathrm{eff}}^2}{i(\omega_m-\omega)+\kappa_m}} 
\label{S11}
\end{equation}
for reflection.
Here, $\kappa_{e}$ are the losses due to the coupling to the microwave feedline into the resonator, $\kappa_c$ the total (loaded) cavity resonator losses, $\omega_c $ is the resonance frequency of the cavity resonator, $\omega_m$ is the frequency of the magnons, and $\kappa_m$ is the loss parameter for the magnons corresponding to the magnon linewidth.

\subsection{Scattering parameters for two-port driven CMPs}

In order to harness the CMP for real applications, it is not sufficient to only obtain a strongly coupled cavity-magnon system, but instead, the coupling strength as a measure for coherent information exchange needs to be controlled. 
Among other ways to achieve a control of the coupling strength (c.f. \cite{Bernier2018,Harder2018a}), the approach of the introduction of a second microwave port to the system represents another possibility to obtain such a control \cite{Grigoryan2018,Boventer2019}. \newline
For this two-port driven CMP, the above assumptions for the derivation of the $S$-parameters remain valid for the cavity port.  However, the second port, which we call magnon port, ideally couples to the magnons only (this is the case shown in Fig.~\ref{Setup} when considering the effect of the green dotted line). 
As a result of the addition of the magnon port, the magnonic subsystem is now directly coupled to the external bath which perturbs the balanced gain and loss of the intracavity system in the presence of the cavity photon coupling without the magnon port \cite{Bender2019}. 
Consequently, the intracavity system describing the cavity photon-magnon coupling is no longer a closed system but an open one.
\newline
Furthermore, the magnon port may differ in phase and amplitude which in addition with the direct coupling to the magnons results in a change of the expression for the scattering parameter $S_{\mathrm{11}}(\omega)$ for a single port driven CMP. This is discussed in the following. 
\\ \newline
As done previously for the simple hybrid system, our approach is based on an interaction Hamiltonian $\mathcal{H}_{\mathrm{sys}}$. However, in order to derive a new expression for $S_{\mathrm{11}}(\omega)$, $H_{\mathrm{sys}}$ is  modified. 
Now, we  assume that the input from the microwave feedline, which couples to the cavity port, is given by $b_{\mathrm{in},1}$. In the same way that the second port, which exhibits the relative phase shift, is given by $b_{\mathrm{in},2}$. The resulting spectrum is recorded at the second port of the VNA which is configured for a transmission measurement. However, the signal there corresponds to the back-reflected signal from the cavity port, given by $b_{\mathrm{out},1}$. 
\begin{figure}[tb]
\centering
\pdfpageattr {/Group << /S /Transparency /I true /CS /DeviceRGB>>}
\includegraphics[scale=0.6]{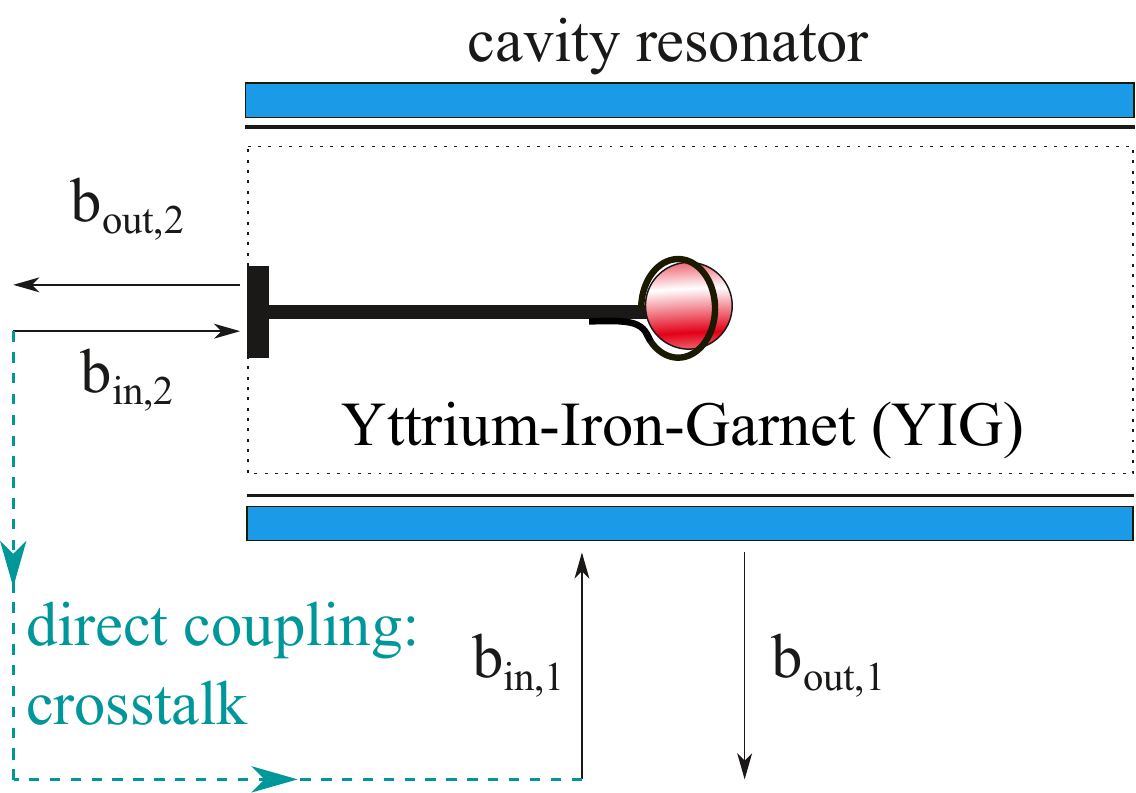}
\caption{Sketch of the different roles of the ports and thus their influence onto the coupled system. The cavity resonator is given as blue horizontal bars while the coupling loop of the second input is shown as a inductive coupler. The magnonic sample (red) is placed at the end.  
Here, $b_{\mathrm{in,1}}$ represents the microwave photon input field from the signal line directly exciting the cavity photons and it is the port where one measures the back-reflected signal and $b_{\mathrm{in,2}}$ is the input signal from the second input with the  additional phase shifter inserted. 
The direct coupling between $b_{\mathrm{in,2}}$ to $b_{\mathrm{in,1}}$ is the what we refer to as crosstalk. 
The fields $b_{\mathrm{out},1}$ and $b_{\mathrm{out},2}$ refer to  the output microwave photon fields of the first and second input, respectively.}
\label{FormulaS11}
\end{figure}
The different roles and the labelling of both feedlines in our systems are sketched in \figref{FormulaS11}. This schematics shows the cavity resonator with the inserted YIG sphere. Considering only $b_{\mathrm{in,1}}$, this input field corresponds to the classical cavity photon magnon-polariton experiments where both subsystems hybridize at resonance and form an avoided level crossing in the dispersion \cite{Tabuchi_2014, Zhang_2014}. The addition of the magnon port (input $b_{\mathrm{in,2}}$) changes the system's properties drastically. 
The system's crosstalk is small for $\delta_0\approx 1$ and is neglected in the following derivation. However, towards higher values of $\delta_0$, it's contribution increases and, hence, lowers the signal to noise ratio and has to be taken into account (c.f. \secref{HigherDelta}).


As previously mentioned, only the microwave photons from the cavity port excite the cavity resonator photons. This excitation is expressed by the photon creation and destruction operators a and $a^\dagger$, respectively. 
Thus, the AC magnetic field originating from the magnon port serves solely as a direct input for the magnons. 
\\\\
The second driving field acts on the magnetisation and, hence, exerts an additional torque on the magnetisation \cite{Grigoryan2018}. If the phase and amplitude of this torque are chosen correctly, this torque compensates all dissipation channels including the coupling of the magnons to the cavity photons, and the avoided level crossing of the CMP coalesces.  As a result, level merging can be observed which also marks the transition to the regime of level attraction.  
\\\\
Therefore, the classical Tavis-Cummings Hamiltonian for a coupled system with N constituents is extended to a \textit{driven} form. As the drive takes place via the coupling of the magnons to the cavity photons, the driving frequency, denoted by $\Omega$, corresponds to the coupling strength $g_{\mathrm{eff}}$ modulated by the relative phase $\phi$ and amplitude $\delta_0$ from the contribution of the second port. It is key that the second port is not just another microwave port of the cavity resonator but acts \textit{indirectly} on the cavity resonator photons via the coupling of the magnons. Otherwise, the effect of a relative phase and amplitude would result in interference effects and not level merging of the CMP's dispersion. 
\\\\
Now, the system Hamiltonian $\mathcal{H}_{\mathrm{sys}}$ has to be modified to take into account this new contribution which results in an open system due to the direct coupling.  
The total number of particles is conserved and, thus, the first two terms denote the total number of cavity photons $\hat{n}^{\mathrm{photons}}=a^\dag a$ and magnons $\hat{n}^{\mathrm{magnons}}=m^\dag m$ in the system. In contrast to \eqnref{SingleToneH}, there are now two interaction terms in the system Hamiltonian. As previously, $H_{\mathrm{int},1}=\hbar g_{\mathrm{eff}}\left(m^{\dag}a+a^{\dag}m\right)$ describes the interaction with coupling strength $g_{\mathrm{eff}}$ of the cavity resonator photons with the magnons and vice versa. The addition of a second interaction term 
$H_{\mathrm{int},2}=\hbar g_{\mathrm{eff}}\delta_0 e^{i\phi}(a^\dagger m)$ considers the impact of the magnon port on the hybrid system via the coupling strength $g_{\mathrm{eff}}$.
\\\\
As a consequence from our open system, we describe the two-port driven CMP by a non-Hermitian Hamiltonian via:
\begin{eqnarray*}
\mathcal{H}_{\mathrm{sys}}=\hbar\omega_{c} a^{\dag}a+\hbar \omega_{m}m^{\dag}m+ \hbar g_{\mathrm{eff}}\left(m^{\dag}a+a^{\dag}m\right) + \hbar \Omega\left(a^{\dag}m\right),
\end{eqnarray*}
where $\Omega=g_{\mathrm{eff}}\delta_0 e^{i\phi}$. 
The last term is now interpreted as an additional drive of the cavity photons through the coupling to the magnons which are excited by magnon port.\\\\
The complex conjugated term of the last term is not included because this would correspond to the crosstalk, the direct interaction between the creation operator of the cavity resonator $a^\dagger$ and the magnon lowering operator $m$. 
The addition of a second microwave input to the hybrid system leads to an additional torque exerted on the precessing magnetisation due to the induced change in the x- and z- components of the AC magnetic fields [c.f. Ref. \cite{Grigoryan2018}]. \\
Depending on the magnitude and the orientation of this torque which is determined by $\delta_0$ and $\phi$, the system's dissipation can be compensated if $\delta_0=1$ or even result in an additional drive for $\delta_0 >1$. In this picture, the coupling strength represents yet another dissipation channel which is then also compensated. Thus,  tuning $\phi$ and $\delta_0$ allows for a control of the coupling strength of the CMP in this specific system. 
\\
However, in order to include a control of the coupling strength via the additional torque which can compensate for the dissipation 
in the system, the above Hamiltonian needs to be non-Hermitian. Also, considering the $2\times 2$ matrix by modeling the CMP, for instance, by two coupled harmonic oscillators  where the off-diagonal elements representing the coupling terms \cite{Harder2018}, level merging  is only possible if the product of the off-diagonal terms is negative. It cannot be a positive, real valued quantity because the interaction potential would be repulsive. Thus, this means that the sign of the off-diagonal product has to change. \\\\
Now, the equations of motion can be written down in Langevin form \cite{Walls2008} as:
\begin{eqnarray*}
\frac{\partial m(t)}{\partial t} &=-i\omega_m m(t)-ig_{\mathrm{eff}}a(t)-\kappa_m m(t)+\sqrt{2\kappa_{e,2}}b_{in,2}(t), \\
\frac{\partial a(t)}{\partial t}&=-i\omega_c a(t)-ig_{\mathrm{eff}}(1+\delta_0 e^{i\phi}) m(t)-\kappa_c a(t)+\sqrt{2\kappa_{e,1}}b_{in,1}(t),
\end{eqnarray*}
where $\omega_m$ denotes the magnon precession frequency, $g_{\mathrm{eff}}$ the effective coupling strength, $\kappa_{e,2}$ the coupling factor to the magnon port, $\omega_c$ the cavity photon frequency, $\kappa_c$ the total resonator losses and $\kappa_{e,1}$ the coupling factor of the cavity port. 
 After a Fourier transformation and employing the Input-Output relation for a system with one external port and a reflection measurement $b_{out,1}+b_{in,1}=\sqrt{2\kappa_{e,1}}a$  (\cite{Walls2008}) the scattering parameter $S_\mathrm{11}(\omega)$ can be expressed as :
 \begin{eqnarray}
 S_\mathrm{11}(\omega)=&-1+\frac{2\kappa_{e,1}}{-i(\omega-\omega_c)+\kappa_c+\frac{g_{\mathrm{eff}}^2(1+\delta_0 e^{i\phi})}{X}}-\nonumber\\ & \frac{2ig_{\mathrm{eff}}\delta_0 e^{i\phi}(1+\delta_0 e^{i\phi})\sqrt{\kappa_{e,1}\kappa_{e,2}}}{{X}\left(-i(\omega-\omega_c)+\kappa_c+\frac{g_{\mathrm{eff}}^2(1+\delta_0 e^{i\phi})}{X})\right)}, 
\label{S11_LM}
 \end{eqnarray}
where $X=-i(\omega-\omega_m)+\kappa_m$ . The first two terms can be mapped to Eq.~(\ref{S11}) except a change in the term for the coupling strength from $g_{\mathrm{eff}}^2 \rightarrow g_{\mathrm{eff}}^2(1+\delta_0e^{i\phi})$. In contrast, the term  considering the coupling strength in the expression for Eq.~(\ref{S11}) for the CMP driven with a single port is purely real. This would be the full expression for the scattering parameter in the case of a single port CMP. However, the additional input via the magnon's coupling to the cavity resonator photons has to be considered for the two-port experiment. Hence, the third term considers this contribution. As the additional drive is mediated by the coupling to the cavity photons, it is proportional to $g_{\mathrm{eff}}$, i.e. the coupling in the limit $\delta_0 \rightarrow 0$ for different phases $\phi$. 

\section{Results and Discussion} \label{Sec:Res}

Having discussed the nature of the hybrid magnon-cavity system under various conditions, we now turn to the direct implications of two ports in a spectroscopic experiment.

\subsection{Two-port spectroscopy numerically analyzed}

We start by looking at the characteristics of Eq.~(\ref{S11_LM}) regarding the coupling strength. One can see that $g_{\mathrm{eff}}$ is completely real for a \akuts{single-port} driven CMP. 
However, in case of a second contribution, the previous expression for the coupling strength has to be rewritten as $g^\prime (\delta_0,\phi)$ which reads as
\begin{equation}
g^{\prime}(\delta_0,\phi)=g_{\mathrm{eff}}\sqrt{1+\delta_0 e^{i\phi}},
\label{FinalGExpr}
\end{equation}
where $g_{\mathrm{eff}}$ corresponds now to the \akuts{single-port} coupling strength, i.e. the coupling strength in the limit for $\delta_0\rightarrow 0$.

\begin{figure*}[tb]
\includegraphics[width=\linewidth]{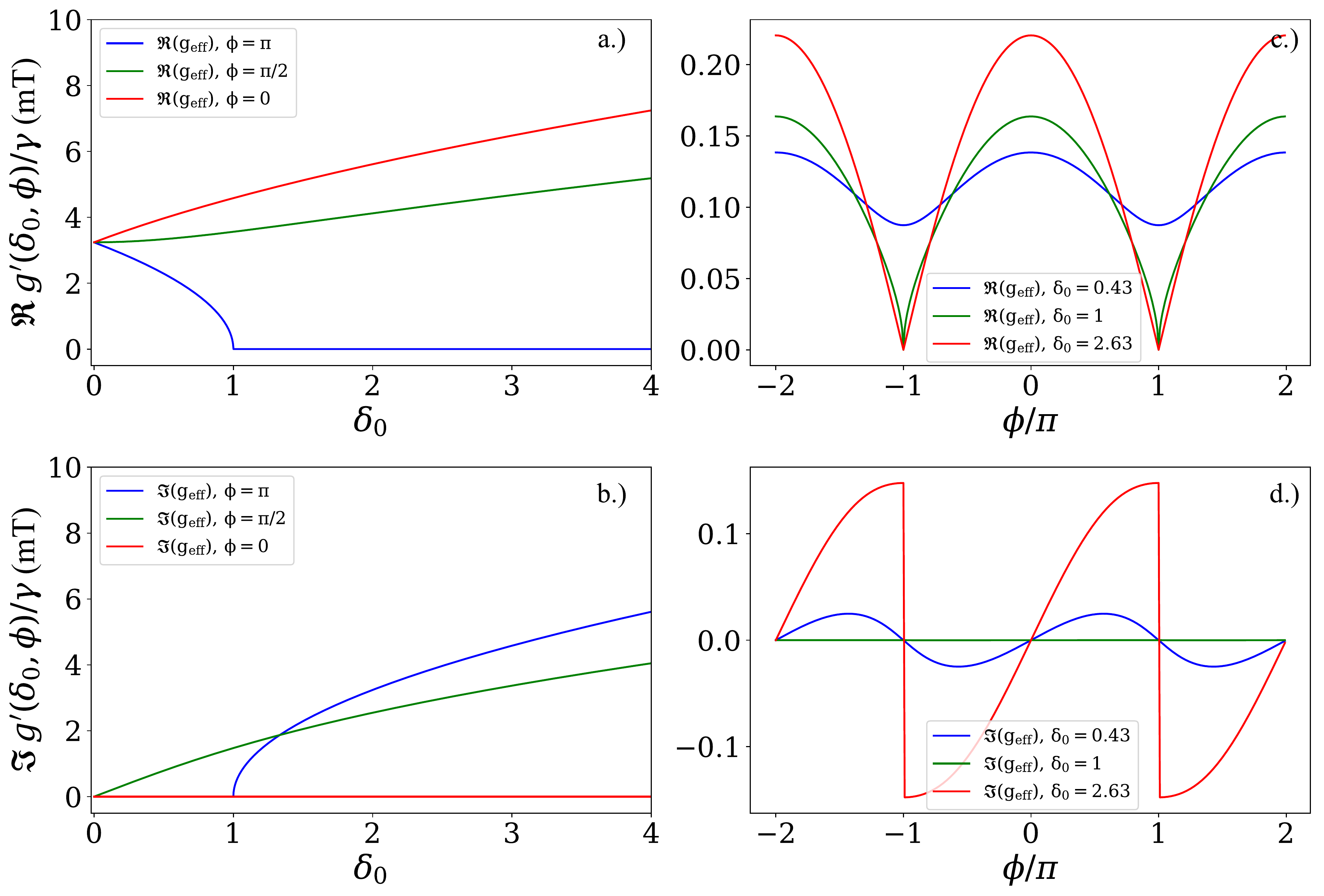}
\caption{Simulations of the dependence of the real and imaginary part of the complex coupling strength on the relative amplitude ratio $\delta_0$ (a.) and b.)) and phase $\phi$. (c.) and d.)). 
a.) Dependence of the real part of the coupling strength for three different values of the relative phase ($\phi \in (0,\pi/2, \pi)$). For $\phi=\pi$, the real part goes to zero for $\delta_0 \geq 1$ whilst for $\phi=0$, the real part continues to increase. 
At the intermediate phase value of $\phi=\pi/2$, the coupling strength also increases but with a smaller gradient compared to $\phi=0$. 
b.) Dependence of the imaginary part of the coupling strength for three phase values. Compared to a.) the imaginary part is always zero for $\phi=0$, non-zero only when $\delta_0\geq 1$ for $\phi=\pi$ and constantly increasing for all values of $\delta_0$ for $\phi=\pi/2$. c.) The real part of the coupling strength as a function of $\phi \in (-2\pi, 2\pi)$ for three values of $\delta_0$ below, at the onset of, and in the regime of level merging. The dependence is periodic for all $\delta_0$, with increasing maxima of the coupling strength for $\phi=0$ and a sharp minimum at $\phi=\pi$ for $\delta_0>1$. Below $\delta_0=1$, the coupling is always suppressed. However, the slope for values close to $\pi$ increases for higher $\delta_0$. 
d.) Imaginary part from the spectrum shown in c.). Above $\delta_0=1$, the plot becomes more and more antisymmetric in the sense of a \akuts{smooth} continuous transition at $\phi=0$ and an increasing discontinuity, i.e sign change, at $\phi=\pi$. }
\label{SimuComplexCoupling}
\end{figure*}

For $\delta_0=1$ and $\phi=\pi$, the term in the square root vanishes and a complete merging of the frequency gap of the avoided level crossing in the dispersion spectrum is expected. 
Hence, this combination of relative phase and amplitude is what we describe as the onset of level merging. 
If the relative phase is kept constant at $\phi=\pi$ and $\delta_0$ is further increased, the term $g^\prime(\delta_0,\phi)$ describing the coupling between the cavity photons and the magnons becomes purely imaginary denoting the regime of level attraction. 
In \figref{SimuComplexCoupling}, the expected dependence of the complex coupling strength on $\delta_0$ [(a) and (b)] and $\phi$ [(c) and  (d)] is displayed for the real [(a) and (c)] and imaginary part [(b) and (d)].
The left column shows the real and imaginary part of the coupling strength as a function of the relative amplitude ratio $\delta_0$ for three fixed values of the relative phase ($\phi \in{0, \pi/2, \pi}$). For $\phi=0$, the coupling strength increases with $\delta_0$ whilst remaining a real valued quantity. On the other hand, for $\phi=\pi$ the real part vanishes for $\delta_0 \geq 1$. Beyond this, the coupling strength is imaginary and increases for higher values of $\delta_0$. A relative amplitude ratio of $\delta_0=1$ constitutes the transition from level repulsion to level attraction via level merging at this specific $\delta_0$ for $\phi=\pi$, because the sign of $g^\prime(\delta_0,\phi)$ in Eq.~(\ref{FinalGExpr}) changes from positive to negative.
Now, if we look back to the framework of two coupled harmonic oscillators, we can see that the repulsion between the anti-symmetric and the symmetric mode is changed to an \akuts{attraction of the eigenvalues} of the coupled system. 
The relative phases of $\phi=0$ and $\phi=\pi$ represent two special cases. Since either the imaginary ($\phi=0$) or the real part ($\phi=\pi$) for $\delta_0=1$ are zero, these cases allow to attribute the real part of the coupling strength to level repulsion and the imaginary part to attraction, respectively.
In this regard, for intermediate relative phase values, the coupling strength is comprised of both a repulsive and attractive contribution. The final shape of the spectrum then depends on whether for a specific relative phase the real or imaginary part is the dominant contribution. 
However, due to the non-zero contribution of the other, the dispersion spectra are slightly distorted by the coexistence of both repulsion and attraction.  
\newline 
In the case of $\phi=\pi/2$ [real part shown in \figref{SimuComplexCoupling}(a) and imaginary part shown in \figref{SimuComplexCoupling}(b)], the non-zero imaginary part acts to \akuts{damp} the increase of the coupling strength towards higher values of $\delta_0$. 
At this relative phase, both contributions are comparable in magnitude. Therefore, compared to the increase (decrease) for $\phi=0$ and $\phi=\pi$ one should expect a strongly suppressed dependence of the coupling strength on $\delta_0$ for $\phi=\pi/2$.
In addition, the relative amplitude ratio can be kept fixed and the coupling strength studied as a function of the relative phase (c.f. \figref{SimuComplexCoupling} c.) and  d.)). The dependence on $\phi$ is illustrated for three different values of $\delta_0$ in \figref{SimuComplexCoupling} (c) for the real part and (d) for the imaginary part.
\newline
The real part of $g^\prime(\delta_0,\phi)$ displays a periodic dependence on the relative phase in the interval $-2\pi$ to $2\pi$. 
For $\delta_0<1$, the coupling strength increases equally for $\phi=0$ as and $\phi=\pi$ for the same value of $\delta_0$. Hence, the coupling is modulated, but for the regime of level merging the relative amplitude ratio $\delta_0$ needs to be altered. 
For instance, if $\delta_0\geq 1$ (green and red solid lines in \figref{SimuComplexCoupling} (c)), the coupling strength at $\phi=0$ increases. However, at $\phi=\pi$, level merging sets in and the real part of $g^\prime(\delta_0,\pi)$ goes to zero. At this point, the difference between the real part of $\delta_0=1$ and $\delta_0=2.63$ is negligible. 
This changes when the contribution from the imaginary part is also considered. For $\delta_0<1$, the coupling strength $g^\prime(\delta_0=\mathrm{const},\phi)$ is a continuous function for $\phi \in (-2\pi,2\pi)$. However, at the transition to level merging, i.e $\delta_0=1$, it becomes discontinuous at $\phi=\pm \pi$. At this point, the value of the imaginary part of the coupling strength is no longer uniquely defined. When the relative amplitude ratio is further increased, the discontinuity increases both in slope and magnitude. Just as in the previous description [(a) and (b)], the magnitude of the imaginary part is zero for all values of $\delta_0$ when $\phi=0$.

\subsection{Two-port cavity magnon-polariton spectroscopy}

In our experiment for two-port cavity magnon-polariton spectroscopy, the relative amplitude ratio $\delta_0$ is defined as the ratio of the AC magnetic field from the magnon port and the cavity port, that is $\delta_0=\frac{\textbf{h}^{\mathrm{AC}}_\mathrm{x,magnon \, port}}{\mathbf{h}^{\mathrm{AC}}_\mathrm{cavity\, port}}$. 
Please note that in the experiment we are not able to directly measure the strength of the internal AC magnetic fields at the position of the sample. However, we can derive $\delta_0$ from calculating an external amplitude ratio $\delta_{\mathrm{ext}}$ which is defined as $\delta_{\mathrm{ext}}=\frac{A_{\mathrm{magnon\, port}}}{A_{\mathrm{cavity \, port}}}$, where $A$ denotes the amplitude of the microwave feedline at either port before it is coupled into the microwave resonator. 
The efficiency of the coupling, i.e. it's quality factor of the microwave signal into the resonator at either port can be determined by performing a {\color{black} \akuts{circle fit}, i.e. fit in the complex plane of the individual reflection measurement from each port \cite{probst} and yields additional factor $\zeta<1$ to the external amplitude ratio for coupling into the cavity resonator}.
Then, $\delta_0$ is calculated via $\delta_0=\zeta \delta_{\mathrm{ext}}$. 
\newline
The cavity port directly drives the cavity photons, i.e. the specific cavity mode. Typically, its amplitude is much higher than the initial amplitude contribution from the magnon port. 
As a result, in order to increase the value of $\delta_0$, the microwave feedline to the cavity port needs to be attenuated. Attenuating the cavity ports amplitude instead of amplifying the amplitude of the microwave signal which enters at the magnon port clearly prevents us from reaching a nonlinear regime for the CMP but also sets an intrinsic limit to our setup due to the presence of noise. 
The further the cavity port is attenuated, the lower the signal to noise ratio of the recorded data as we probe our system in reflection at the cavity port. Hence, the data analysis is more and more aggravated until clear statements on the specific nature of the signal are not possible any more. The subtle nature of crosstalk from magnon to cavity leading to an increasing signal, whereas the cavity reflection shows up as a decrease from the baselines signal renders the measured response very sensitive to the achievable cross-talk suppression. As for all microwave devices, reduction of unwanted signal leakage is far from trivial. As an example, a crosstalk of 1\% corresponds to -20dB of applied power. A power ratio of -20dB corresponds to an amplitude ratio of 0.1. In this work, the relative signal amplitudes are described by $\delta_0$. That means for $\delta_0 > (0.1)^{-1} =10$ (i.e. +20dB relative power to the magnon compared to the power at the cavity the crosstalk signal from the magnon port dominates the cavity probing signal. 
{\color{black}
\subsection{Mechanism for level attraction for a two-drive CMP system}
\begin{figure}
\centering
\includegraphics[width=\columnwidth]{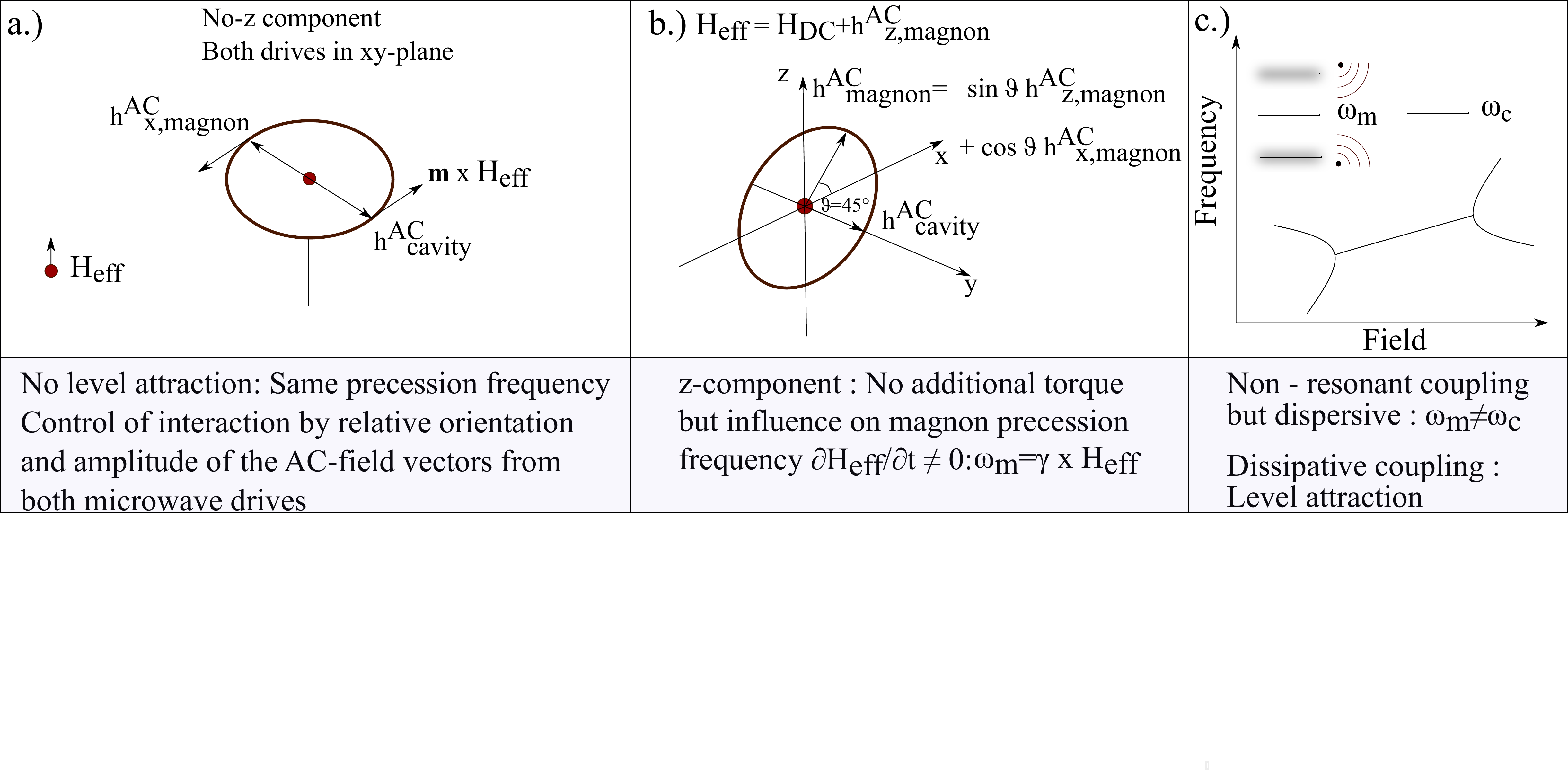}
\caption{Model for the occurence of level attraction in our two-drive controlled CMP. Instead of an impact on the photon frequency, the modulation is via the magnon's frequency. In line with other works (e.g. Ref.\cite{Yao2019a}), the physical mechanism behind our observation of level attraction is also the transition from the coherent/strong coupling regime to the dispersive coupling regime. (a) Situation of  a two tone driven system in the xy-plane only. There is no level attraction possible because the only interaction is done by the superposition of both in-plane components. 
(b) The magnon port is tilt by 45°, which yields a  time dependent AC component in the z-Plane and modulates the effective magnetic field.
(c) Illustration of the change of the magnon frequency which results in a deviation from the coherent coupling ($\omega_c\equiv \omega_m$) regime and allows for the observation of level attraction (similar to Ref.\cite{Yao2019a} for a CMP driven by one microwave input port). }
\label{Explanation}
\end{figure}
To date, level attraction in CMP systems has been experimentally observed by different approaches employing a single input (e.g. Ref. \cite{Harder2018a}, \cite{Bhoi2019}). The microscopic origin of level attraction is now explained by the dominance of dissipative coupling over the typically much stronger coherent coupling between cavity photons and magnons \cite{Yao2019a}. The hallmark of the coherent coupling regime is the occurrence of an avoided level crossing. 
For the explanation of the mechanism in our two-microwave drive experimental apparatus, we follow the microscopic model presented in Ref. \cite{Yao2019a} of either coupling to a standing wave (coherent coupling regime with $\omega_m \equiv \omega_c$) or travelling wave ($\omega_c \neq \omega_m,\, \omega_m=\mathrm{const.}$). 
Although our approach rather addresses the magnon frequency instead of the cavity frequency, we employ the same physical mechanism of a transition between the coherent to the dissipative coupling regime which allows us to also observe level attraction. As can be inferred from Fig. 1b, the coupling loop which denotes the magnon port, exhibits an angle of $45^\circ$ to the xy-plane. In our system, it is that non-zero contribution $h^{\mathrm{AC}}_{\mathrm{magnon}}$ parallel to the effective magnetic field $H_{\mathrm{eff}}$ and, hence, the saturation magnetization, which results in the possibility to observe level attraction by appropriately tuning the relative phase and amplitude if both drives. ß
If $h^{\mathrm{AC}}_{\mathrm{magnon}}=0$ (c.f. Fig.\ref{Explanation} (a), the coherent exchange of energy can be controlled by appropriately superimposing the contributions from either drive but now level attraction is observed. 
However, if $h^{\mathrm{AC}}_{\mathrm{magnon}} \neq 0$ (c.f. Fig.\ref{Explanation}) (b)), the frequency of the magnetization precession, i.e. the magnon frequency, is modulated by time dependent addition of the AC magnetic field to $H_{\mathrm{eff}}$. Similar to Ref. \cite{Yao2019a}, that change results in a \akuts{detuning} of the magnon from the cavity photon in terms of frequency. As a result, the system is less coherently coupled and the contribution of the dissipative coupling increases. 
In our system, level attraction is observed for $\phi=\pi$ and $\delta_0>1$, such that the coherent coupling is suppressed and the detuned, i.e. dissipative contribution is dominating (c.f. Fig. \ref{Explanation} (c)).
}
\subsection{Interplay of attraction and repulsion for intermediate phases ($\delta_0>1$)}
\begin{figure}[tb]
    \centering
    \includegraphics[width=\columnwidth]{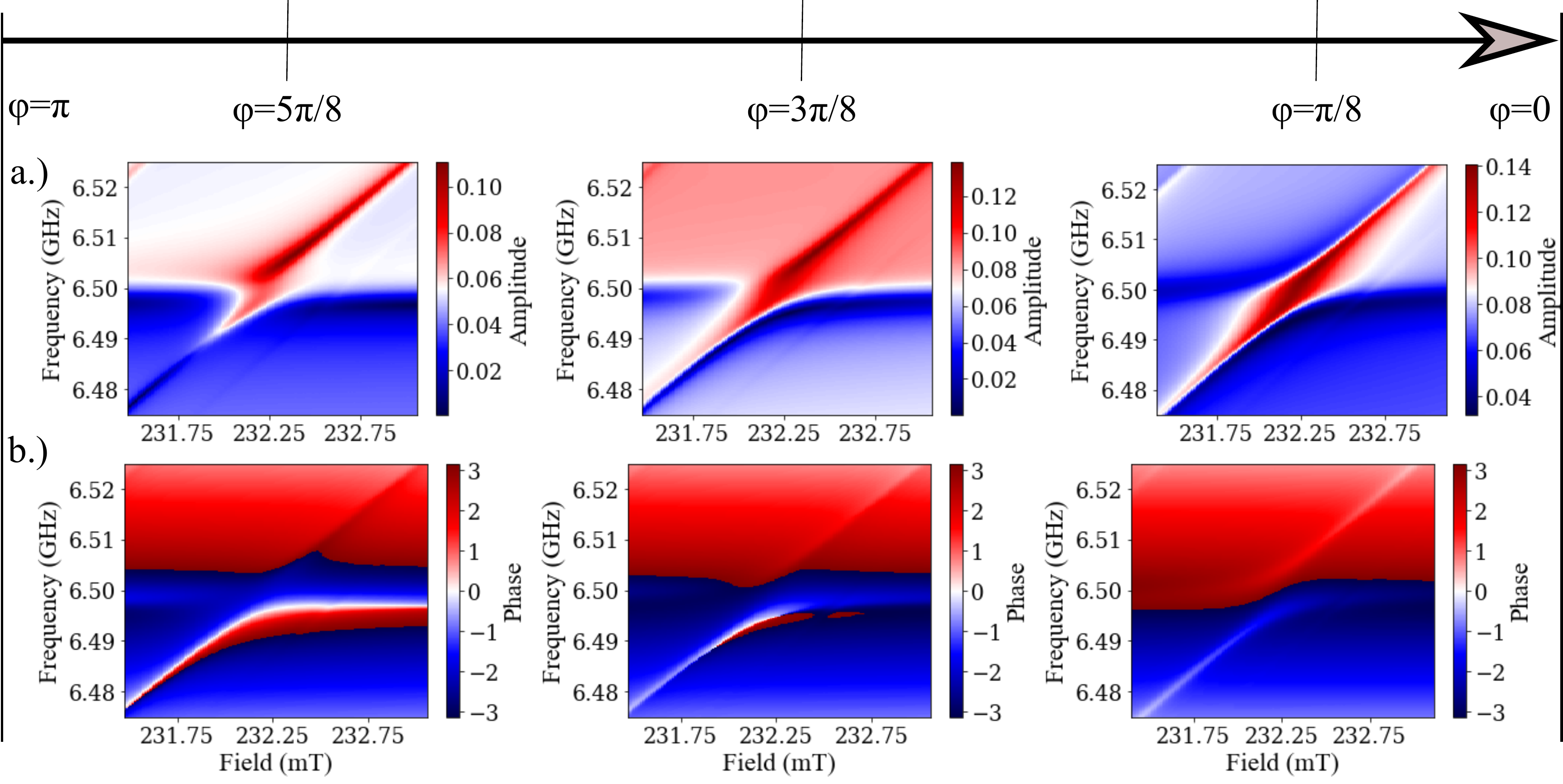}
    \caption{Experimental data showing the coexistence of level repulsion and attraction for $\delta_0=1.31\pm 0.22$ and for different values of intermediate phases both for amplitude (a.) and phase (b.) {\color{black}at the transition from level attraction (left) towards level repulsion (right)}. The counteracting repulsion and attraction at resonance lead to a partial extinction and partial enhancement of the signal. One can see both the characteristic features. First, one can infer the signal's curvature corresponding to the symmetric and antisymmetric mode of a \akuts{classical} avoided level crossing. Second, the existence of the  {\color{black}level attraction structure with two triangles below and above the magnetic field for a resonant coupling (i.e. here a crossing of the magnon and cavity photon dispersions -- c.f. also the phase signal) with the right apex more dominant than the left one is visible. The dominance of either phase depends on the chosen intermediate phase value and the relative distance to a phase of $0$ or $\phi$. } }
    \label{Coexistence}
\end{figure}

For the intermediate phases, we observe a coexistence of level merging and level attraction. 
{\color{black} Specifically, in \figref{Coexistence}, we show the coexistence of level attraction ($\phi=\pi$) and level repulsion ($\phi=0$) for a series of phase values between $0$ and $\pi$ for $\delta_0=1.31\pm 0.22$. Depending on the relative difference of the actual relative phase value, the spectrum exhibits a stronger contribution from either level attraction (e.f. left column in \figref{Coexistence}) or level repulsion (e.g. right column in \figref{Coexistence}). For instance, for the middle column ($\phi=3\pi/8$), below resonance (frequencies below the frequency of the cavity photon $\omega_c=6.5\,\mathrm{GHz}$), the signature of an avoided crossing with a beginning opening of an anticrossing gap is visible. However, above resonance (frequencies above the frequency of the cavity photon $\omega_c=6.5\,\mathrm{GHz}$) partially the triangular shape of the level attraction regime (c.f. \figref{Explanation} c.)) is also visible, showing a almost equal contribution of both coupling regimes.
As it is clearer to see this difference in the spectrum showing the phase of the coupled system (\figref{Coexistence} b.)), we also plot the phase. As can be seen from the clear phase jump and corresponding anticrossing in the right spectrum of \figref{Coexistence} b.) the contribution of level attraction is negligible to the complete system's response which is in stark contrast to the phase spectrum for $\phi=5\pi/8$ where the typical shape of a spectrum for level attraction (such as sketched in \figref{Explanation} c.)) can be seen.  Thus, we are able to deliberately tune the relative contributions from level attraction and level repulsion to the total signal with our specific system. }
Apart from showing the broad tunability of our two-port driven approach to control the cavity magnon-polariton, the control of the relative contribution from level attraction  (level merging) and repulsion (anticrossing) might be interesting to generate intermediate states between maximum or minimum entanglement of cavity photon - magnon states and, hence, the transfer of information.
For instance, the two-port driven CMP can be transferred to the millikelvin temperature regime and these concepts tested in the single magnon regime as proposed in Ref. \cite{Yuan2019}.

\subsection{Towards high values of $\delta_0$}
\label{HigherDelta}
\begin{figure}[tb]
\centering
\includegraphics[scale=0.5]{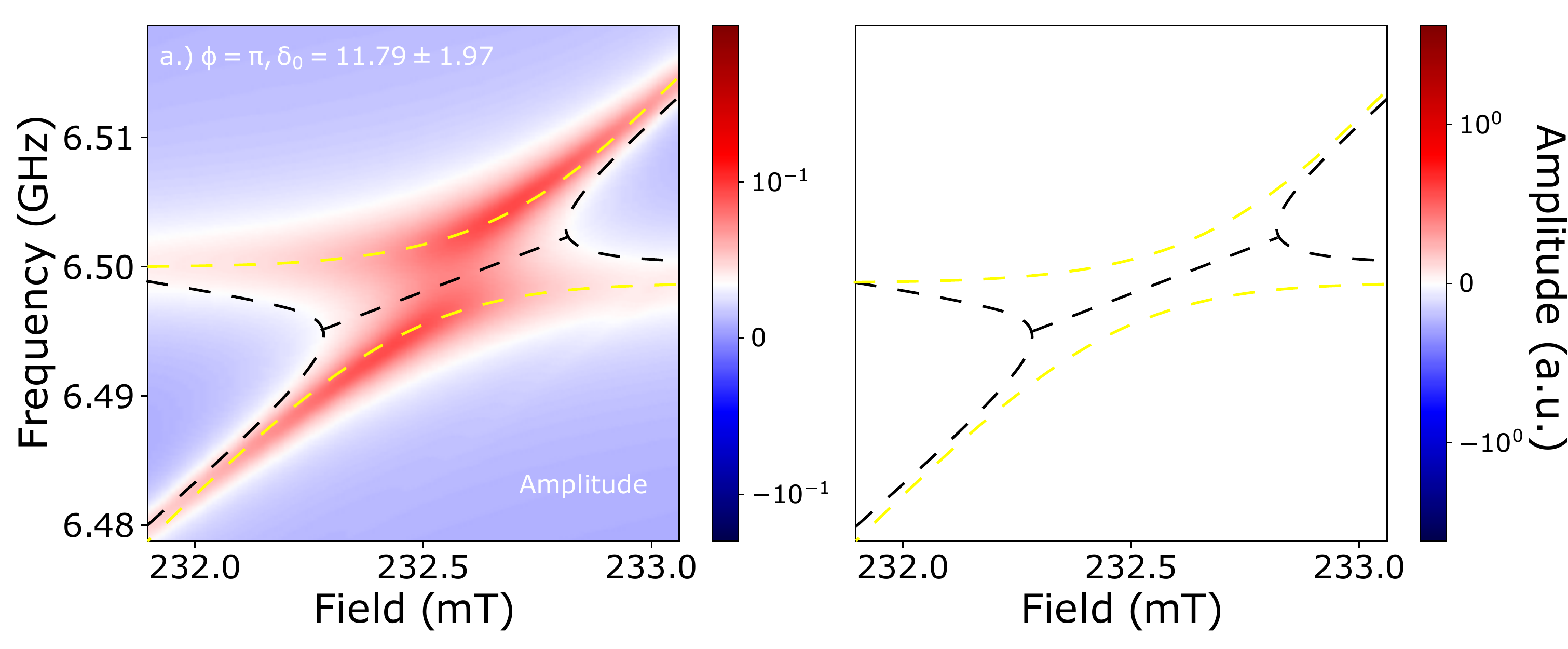}
\caption{Dispersion spectra of the amplitude (a.)) and phase (b.)) for $\phi=\pi$ and highest measured value of $\delta_0=11.79\pm 1.97$ in a logarithmic scale. The spectra are a superposition of two signals, as indicated by the dashed lines, which serve as a guide for the eye. They are comprised of a level merging spectrum (black) with an additional avoided level crossing (yellow) at the same resonance frequency. 
The relative weight of the crosstalk measured in transmission at the cavity port increases towards higher $\delta_0$ and has to be taken into account. 
Thus, for the complex-valued coupling strength, this avoided level crossing adds a parasitic real-valued contribution, which decreases the field distance between the apexes of the level merging signal and has to be considered in the calculation of $\Im(g^\prime(\delta_0,\phi))$}
\label{Cavity_Att}
\end{figure}
As shown in Fig. \ref{Cavity_Att} (a) for the amplitude and (b) for the phase response for $\phi=\pi$, for our system,  the highest value was found to be $\delta_0=11.79\pm 1.97$. 
The dashed lines serve as a guide for the eye and denote the level merging (black) and an anticrossing (yellow) spectrum. 
Whilst the first is the signal of interest, the latter is a result from a direct crosstalk, of our system which was suppressed as much as possible in the experiment but still non-zero [c.f. Ref.\cite{Boventer2019}]. 
Ideally, the AC field contribution from the magnon port does not couple to the cavity port. However, in case of a direct coupling, i.e. crosstalk, the magnon port serves as the input port and we measure an additional transmission signal at the cavity port due to that crosstalk. At the conditions for resonant coupling, the usual hybridization of a single-port driven CMP sets in and is observed by an anticrossing. 
Thus, we measure the superposition of our level merging spectrum and the anticrossing due to crosstalk. 
An attenuation of the cavity port results in an increasing contribution of the magnon port which starts to dominate for $\delta_0 > 1$.
Hence, for high values of $\delta_0$, the transmission signal due to crosstalk is higher in amplitude than the reflection signal of interest from level merging (c.f. \figref{Cavity_Att} (a.)). 
\newline
Consequently, for higher values of $\delta_0$ where the exact value of $\delta_0$ depends on the intrinsic amount of suppressing the crosstalk, it is not sufficient to only take the amplitude data into account to clearly identify the presence of level attraction of our system. Therefore, the phase data has to be considered as well. As shown in \figref{Cavity_Att} (b.)) and indicated again by the dashed black (level merging) and yellow (crosstalk anticrossing) it confirms the level merging signal for $\delta_0=11.79\pm 1.97$.  {\color{black} It shows, that the width in terms of applied magnetic field values around the resonance magnetic field ($H_{\mathrm{res}}\approx 233.6\,\mathrm{mT}$) where the coupled system exhibits a coalesced spectrum can be increased from zero at level merging, i.e. a direct crossing of the cavity photon and the Kittel mode dispersion curves for $\phi=\pi$ and $\delta_0=1$ \cite{Boventer2019}, to $\approx 0.5 \,\mathrm{mT}$ by altering the value of $\delta_0$ towards higher values. Simulations with \eqnref{S11_LM} also show that this distance can be further increased for even higher values of $\delta_0$ but due to the increasing contribution of crosstalk, our specific system is meeting its experimental limits for $\delta_0=11.79\pm 1.97$.} 
\sectionmark{Summary and outlook}

\section{Summary and outlook}
$ $
In summary, we explained in detail one experimental approach to control the coupling strength by employing the relative phase and amplitude ratio $\delta_0$ of a two-port driven CMP. 
We numerically studied our new expression for the regime of level merging with complex coupling strength. 
Furthermore, we experimentally demonstrated the coexistence of level attraction and level repulsion and the characteristics of the two-port driven CMP in the limit of high $\delta_0$. 
Such coexistence not only demonstrates the broad tunability of our approach, but also how it is possible to realise a type of \akuts{superposition} states of the avoided level crossing and level merging regime where the amount of transmitted information flow can be exactly set.
Since increasing $\delta_0$ results in an enhancement of the relative weight of the crosstalk in the recorded signal, i.e. lowers the signal-to-noise ratio, we also show limitations of controlling the two-port driven CMP's coupling strength.

Moreover, we show that the system's Hamiltonian is non-Hermitian but depends on the phase and amplitude configurations and it can still result in real eigenvalues of the CMP. 
This can be possible because the introduction of a non-Hermitian term into the Hamiltonian denotes the possibility for an open system, i.e. dissipation is now included which is also referred to as approximate non-Hermiticity \cite{Bender2002, Bender2018}. 
For instance, this also describes radioactive decay or the introduction of dissipative systems in semiconductor physics. 
However, even for non-Hermitian systems, the spectra can be real if the system is $\mathcal{PT}$ symmetric, i.e. is invariant under parity and time reversal transformations such that $[H,\mathcal{PT}]=0$.  
$\mathcal{PT}$ symmetric systems are studied in many different fields such as in quantum mechanics \cite{Bender1998}, optical microcavities \cite{Wen2018} or magnetism and magnonics \cite{Galda2018}. This symmetry also started to receive interest in cavity spintronics and for CMPs where the spectra and behaviour of $\mathcal{PT}$ symmetric CMPs have recently been discussed \cite{Zhang2017a,MHarder2017, Cao2019}.  

As shown in Ref. \cite{Zhang2017a}, the $\mathcal{PT}$ symmetric state is achieved by carefully engineering the losses from the cavity resonator and the magnons such that $\gamma_a=\kappa_c=\kappa_m$. 
Then, the coupling strength is tuned by moving the position of the YIG sphere in the cavity resonator. In case of $g_{\mathrm{eff}}=\gamma_a$, the two separate eigenmodes of the coupled system coalesce to one point. This singularity in the eigenvalues represents the hallmark of a non-Hermitian system and this point is called an exceptional point (EP). What we show here, is the possibility to transition from avoided level crossing to level merging by tuning the relative orientation and amplitude of the additional torque added to the system. However, neither the cavity dissipation nor the magnon dissipation are directly accessed and tuned such as has been done in Ref. \cite{Bernier2018}. Rather, we change the relative contribution and orientation of the additional torque, which then enhances or compensates the intrinsic system's dissipation. {\color{black}As a result, in addition to the high tunability between different coupling regimes, our two-port driven approach offers the possibility for further studies towards $\mathcal{PT}$ symmetric magnon polaritons. However,}
the connection and incorporation of the experimental results from this two-port driven system to the above discussion of $\mathcal{PT}$ symmetry and singularities such as EPs and requires further in-depth theoretical studies.

Finally, here we demonstrate control over the coupling regime without any direct changes of the experimental setup, thus improving measurement and analysis precision and being advantageous for real applications. Such control mechanism over the spin-photon interaction could pave the way for deliberately turning on and off the coherent exchange of information. That could enable future applications for data storage and information processing. {\color{black} For instance, the addition of a non-linear component such as a superconducting circuit to the spin-photon system and the control over the coupling strength could control the photon mediated interaction between the superconducting circuit (processing unit) and the magnons (storage unit). Furthermore, 
by performing fast manipulations of the polariton modes with two independent but coherent pulses to the cavity and magnon system \cite{Wolz2019} building blocks for a quantum internet can be realized, and thus, pave the way for further magnon-based quantum computing research.}

\section*{Acknowledgement}
We acknowledge valuable discussions with Dmytro Bozhko, Tim Wolz, Can-Ming Hu, Bimu Yao, Vahram L. Grigoryan, Ka Shen, and Ke Xia. This work is supported by the European Research Council (ERC) under the Grant Agreement 648011 and the DFG through SFB TRR 173/Spin+X. R. Mac{\^e}do acknowledges the support of the Leverhulme Trust. 

\providecommand{\newblock}{}

\end{document}